# Microstructure-dependent local strain behavior in polycrystals through *in situ* scanning electron microscope tensile experiments

M.A. Tschopp[a,b], B.B. Bartha[c], W.J. Porter[a,d], P.T. Murray[d], S.B. Fairchild[a]

[a] Air Force Research Laboratory, Materials and Manufacturing Directorate, AFRL/RX
Wright Patterson AFB, OH 45433

[b] Universal Technology Corporation
Dayton, OH 45433

[c] United Space Alliance
Cape Canaveral AFB, FL 32815

[d] University of Dayton Research Institute
Dayton, OH 45431

Digital image correlation of laser-ablated platinum nanoparticles on the surface of a polycrystalline metal (nickel-based superalloy René 88DT) was used to obtain the local strain behavior from an *in situ* scanning electron microscope tensile experiment at room temperature. By fusing this information with crystallographic orientations from EBSD, a subsequent analysis shows that the average maximum shear strain tends to increase with increasing Schmid factor. Additionally, the range of the extreme values for the maximum shear strain also increases closer to the grain boundary, signifying that grain boundaries and triple junctions accumulate plasticity at strains just beyond yield in polycrystalline René 88DT. In situ experiments illuminating microstructure-property relationships of this ilk may be important for understanding damage nucleation in polycrystalline metals at high temperatures.

Nickel-based superalloy; in situ SEM; Structure-property relations; Plasticity

I.     INTRODUCTION

Fatigue variability plays a crucial role in determining the total life of fracture critical turbine engine components. The Air Force Research Laboratory's Engine Rotor Life Extension and Materials Damage Prognosis programs have examined life-limiting factors in an effort to extend the lives of service components. Various material-specific mechanisms contribute to fatigue variability (1). For example, fatigue variability has been associated with competing failure mechanisms in Ti-6246, René 88DT and IN-100 (2-4). The ability to predict fatigue variability and the minimum life of critical components significantly affects the sustainability of an aircraft fleet.

To accurately predict fatigue variability, it is vital to understand what the fatigue-critical microstructural features are and exactly how these may couple with loads and temperatures to nucleate damage in these critical components. It is a commonly held notion that damage nucleates in locations of large strain concentrations or where substantial inelastic deformation exists. Digital image correlation (5-9) is a technique often used at larger scales to investigate how strain localizes around part geometric features, such as cracks, holes, and notches. Recently, this technique has been applied at increasingly smaller scales, i.e., *in situ* scanning electron microscope (SEM) (10-14) and atomic force microscope (15-18) studies are now on the order of the underlying microstructural features. For example, Sutton et al. (10, 11) used SEM imaging with digital image correlation (DIC) to validate the experimental strain measurement capability over a large range of magnifications. Additionally, Chasiotis and Knauss (16) combined DIC with AFM images to measure the elastic tensile properties of surface-micromachined polysilicon specimens for MEMS applications. More recently, Xu et al. (19) have combined DIC with AFM images to map the nanoscale wear field of a gold coating. In order to obtain strains at these scales, it is important to have sufficient contrast in the images to perform DIC, i.e., a speckle pattern in SEM imaging applications or a contrast in topography for AFM imaging applications.

*In situ* SEM studies are also used to understand how microstructure evolves with deformation (20-27). For example, Boehlert and colleagues (20) studied the microstructural evolution of TiB whisker microcracking of boron-modified Ti-6Al-4V using in situ SEM tensile experiments at room temperature and 480°C. In polycrystalline materials, researchers have combined in situ SEM experiments with electron backscatter diffraction (EBSD) to assess, for example, grain boundary migration during recrystallization (22-25), phase transformations (26), and crystallographic orientation evolution of single crystals during tensile deformation (27). The ability to combine these studies and understand how specific microstructure features evolve and couple with local strains can greatly enhance our ability to predict fatigue and, perhaps, engineer better materials for fatigue.

This paper presents an *in situ* SEM technique that can be used to obtain the local deformation behavior of polycrystalline materials at room and elevated temperatures. René 88DT, a forged polycrystalline Ni-based superalloy used in aircraft engine components, was chosen as a novel material for this work. The results correlate the local strain behavior obtained from DIC to grain boundaries and grain orientations using EBSD. Ultimately, the objective of this research is to understand how the microstructural variability of polycrystalline materials influences fatigue variability. This paper highlights a methodology that couples local strains with crystallographic orientations to analyze microstructural factors that may influence damage nucleation and fatigue.

## II. EXPERIMENTAL APPROACH

The experimental setup consisted of a screw-driven 1000 lb capacity tensile stage (Ernest F. Fullam, Inc.) placed inside the chamber of a Quanta 600 FEG SEM. The tensile specimen was a flat dog bone-shaped specimen with gage dimensions of approximately 2.80 mm wide x 0.72 mm thick x 10 mm long.

Digital image correlation often uses a speckle pattern to track displacements. In this study, the specimen surface was coated with Pt nanoparticles (Figure 1) using a laser ablation process termed **through thin film ablation** (TTFA) (28). For optimal imaging and deposition, the specimen surfaces were first mechanically polished to a 1 μm finish. The TTFA technique used a 10 nm Pt thin film deposited onto a fused silica plate transparent to the laser wavelength (wavelength = 248 nm, energy density = 0.5 J/cm$^2$). The chamber was filled with Ar at a pressure of 5 torr. The laser irradiates the Pt thin film from the backside, propelling Pt nanoparticles at the intended target, i.e., in this case, the tensile specimen. The density of the nanoparticles on the pattern surface can be controlled by the number of laser pulses. The high density of the Pt nanoparticles compared to the René 88DT provided sufficient contrast in the secondary electron images for DIC. Figure 1 shows the speckle pattern obtained from this technique along with a close up of the deposited nanoparticles. Notice that the multiple laser pulses required with this technique results in some adjacent nanoparticles. The TTFA nanoparticles are deposited in a random speckle pattern that allows sub-pixel resolution of displacement. Platinum nanoparticles enable DIC at higher temperatures than Au patterns (e.g., Ref. 10), i.e., into the temperature range where nickel-based superalloys are typically used. Developing methods for creating random speckle patterns on the specimen surface is still an area of active research for digital image correlation at small scales; other techniques, such as spin-casting nanoparticles (29-30) and lithography-based techniques (31) have also been proven useful for DIC applications. Other experimental methods for measuring local strains exist, i.e., alternative approaches to measuring local strains is through the use of microgrid techniques (cf. 32, 33) or even through electrochemically etching the sample to provide sufficient surface contrast (34-36).

The specimen was then mounted into the tensile stage grip fixture and preloaded to help align the specimen. Both secondary electron (SE) and backscatter electron (BSE) images were collected at 16 bit depth at a pixel resolution of 4096 by 3773 pixels. The SE images were used for digital image correlation while the BSE images provided sufficient detail for aligning the in-situ images with post-processed EBSD images. At each load step, the stage controller was turned off while acquiring the images to minimize any potential distortion effects due to the motor operating the stage. Each image was focused by adjusting the stage fixture to keep the same working distance; this minimizes any additional artifacts due to focusing with the beam only.

The tensile specimen was loaded in 12 steps to 1381 MPa (yield regime) with larger (smaller) step sizes in the elastic (plastic) regime. The present analysis was conducted at room temperature. After unloading the specimen, fiducial marks were used to mark the region of interest, the Pt nanoparticles were removed from the surface and EBSD was used to measure the crystallographic orientations of the underlying microstructure.

Digital Image Correlation was then performed on each image with ViC-2D (Correlated Solutions, Inc.) to calculate the displacement field for each load increment in the 274 μm by 230 μm region of interest. Digital Image Correlation is an optical method to measure deformation on an object surface through tracking the intensity value pattern in small neighborhoods (subsets) during deformation. The image at the 0 MPa unloaded condition was used as a reference image for the displacement calculation and the deformed image was captured at a nominal stress of 1280 MPa. A subset size of 99x99 pixels (~7x7 microns) was used for each calculation with a 5 pixel step size. In other words, every 5 pixels in the reference image, a 99x99 pixel local region in the reference image is located in the deformed image through optimizing the correlation between the local regions (subsets). A cubic B-spline was used to interpolate non-integer intensity values to obtain subpixel accuracies in the displacement calculations. The displacement maps were then converted to maximum shear strain maps for the remainder of the analysis.

## III. RESULTS

Figure 2 shows the result of digital image correlation for the deformed image with a nominal stress of 1280 MPa. Fig. 2(a) shows the reference image, while Fig. 2(b) shows the deformed image with evidence of slip bands in some grains; the loading axis is horizontal. The digital image correlation software uses Fourier transforms of multiple subsets of these images to calculate the sub-pixel displacements and strains, as shown in Fig. 2(c). The average strain in the loading axis direction for the 274 µm by 230 µm region of interest is 0.02. The high strain concentrations are localized in bands oriented approximately 45 degrees from the tensile direction, as would be expected. The bands present in Fig. 2(c) do not necessarily correlate with the observed slip bands in Fig. 2(b); grain boundaries also play an important role in the high local maximum shear strain bands.

Further analysis requires the maximum shear strain maps to overlap the crystallographic information obtained from the EBSD scan. By aligning the inverse pole figure in Figure 3(a) with the BSE image in Figure 3(b), the strain maps can be correlated with information obtained from crystal orientations, e.g., Schmid factor, Taylor factor, etc. The image quality (IQ) map (Fig. 3c) is a quantitative measure of the fit of the Kikuchi pattern from the EBSD scan. The grain boundaries have a lower IQ value than the grain interiors, which allows for a better alignment with the BSE image. The IQ map image is then aligned with the BSE image through a series of steps. First, control points were selected in both images at triple junctions. Triple junctions with a high degree of contrast in surrounding grains for the BSE image were selected since these were easily resolved in the IQ map image. A least squares linear conformal transformation (no distortion) was then used to translate, rotate, and rescale the IQ image into alignment with the BSE image, as shown in Figure 3(d). However, the pixels and their spacing were still different between the two datasets. Therefore, a nearest neighbor interpolation is used to match the pixels in the IQ map with the pixels in the BSE image. All subsequent analyses are related to correlating the shear strain behavior in Figure 2 with quantitative information relating to the crystallographic grain orientation.

Figure 4 shows the correlation between the Schmid factor and the maximum shear strain. Each data point represents 1 pixel from the strain map of Figure 2, i.e., over 460,000 data points total. Figure 4(a) shows the distribution of Schmid factors within the René 88DT grain structure over the same area as Fig. 2(c). The Schmid factor resolves the tensile stress onto the {111} slip plane in the <110> slip direction, i.e., a higher Schmid factor should coincide with a higher shear stress in the direction of slip. Fig. 4(b) shows the relation between the maximum shear strain and the Schmid factor. At low Schmid factors, the range of the extreme values of maximum shear strain (i.e., the low and high values) is not as large as for higher Schmid factors. There appears to be no decisive relationship between the Schmid factor and extreme values of the maximum shear strain, since high maximum shear strains are observed in regions with Schmid factors of approximately 0.35. The high extreme values of maximum shear strain are of particular interest because they indicate regions of damage accumulation, which is important under fatigue conditions. The average maximum shear strain also shows a net increasing trend in maximum shear strain with increasing Schmid factor, although this trend is minimal.

The grain boundary network may also be associated with the localization in strain within the region of interest. Figure 5 shows the correlation between the distance from the grain boundary and the maximum shear strain. The grain boundary pixels were identified by determining if there were two or more grains present in adjacent pixels (4-neighborhood). Fig. 5(a) shows the maximum shear strain map with the grain boundary pixels in black. The distance from the boundary was calculated using a Euclidean distance transform. Interestingly, Fig. 5(b) shows that the upper (lower) extreme values for maximum shear strain decreases (increases) as the distance from the grain boundary increases. This indicates that the grain boundary has a greater propensity to accommodate strain than the grain interiors --- both lower and higher shear strains. The strain behavior at large distances (approximately 8 microns and greater) from the boundary is related to the few large grains within the region of interest. The strain behavior at intermediate distances from the GB encompasses a large number of grains, yet the strain range is not as large as at the boundary. The average shear strain behavior is equivalent for the first 6 μm from the boundary; the deviation at larger distances is affected by the few large grains. These results indicate that shear strains are much more likely to localize at high values at the grain boundary regions rather than the grain interiors.

These trends may not appear as strong because the strain localization after yield is not merely a function of the crystallographic orientation of the underlying lattice, but may also depend on the grain boundary structure, triple junctions, the grain size, and the neighboring grains, *i.e.*, due to multiple factor interactions. In this analysis, the trends apparent in Figs. 4 and 5 indicate that the factors examined in this paper may be significant, though. Additionally, the maximum shear strains and grain orientations obtained are also a two-dimensional representation of a three-dimensional problem, which may further complicate correlating strains with the underlying microstructure. Ideally, having a specimen thickness on the order of the grain size may better elucidate some of these trends. In this work, however, the properties of interest correspond to Rene 88DT grain sizes analyzed. Last, OIM maps *after* deformation were used to supply the grain orientation information. Future work focuses on modifying this methodology to obtain EBSD crystallographic orientation *before* the in situ tensile experiment as well. This is important for high uniaxial strains, which could lead to grain rotation and grain boundary sliding at high temperatures.

## IV. CONCLUSION

In summary, this paper presents a methodology for preparing tensile specimens for *in situ* SEM digital image correlation through a laser ablation process, through thin film ablation. By combining deformation strain maps from DIC with EBSD data, the correlation between the maximum shear strain and a number of microstructure-dependent parameters can be ascertained, e.g., Schmid factor (Fig. 4) and distance from grain boundary (Fig. 5) in this analysis. On average, the maximum shear strain tends to increase with increasing Schmid factor. The range of the extreme values for the maximum shear strain also increases closer to the grain boundary, signifying that grain boundaries and triple junctions accumulate plasticity at strains just beyond yield within polycrystalline René 88DT.

This analysis shows that the strain localization in polycrystalline superalloys, which is important to plasticity, fatigue, and fracture, is a combination of a number of factors related to grain orientation and the grain boundary network. This will require coupling between further experiments and computational approaches to fully understand, and is vital to understanding how damage nucleates in fatigue-critical polycrystalline components. Furthermore, results of this ilk may also be used to estimate constitutive parameters with inverse computational methods based on full-field measurements (cf. 37-38). Future work aims to extend this approach to higher temperatures.

This work was performed at the Air Force Research Laboratory, Materials and Manufacturing Directorate, AFRL/RXLMN, Wright-Patterson Air Force Base, OH. The financial support of the Air Force Office of Scientific Research (AFOSR) through the AFOSR task #92ML02COR with Dr. Victor Giurgiutiu as the program manager is gratefully acknowledged. The authors acknowledge Bob Wheeler and the Material Characterization Facility at the AFRL for their help in this work.

**Figure Captions**

**Figure 1.**  BSE images of nanoparticles on the surface of the tensile specimen.

**Figure 2.**  (a,b) SE images of René 88DT with nanoparticles (bright spots throughout image) at no load and after yield. (c) Digital image correlation was used to generate the maximum shear strains within the 274 μm by 230 μm region of interest.

**Figure 3.**  Image alignment process: (a) Inverse pole figure showing crystal orientations from EBSD scan, (b) BSE image of René 88DT prior to deformation, (c) IQ map of microstructure, and (d) aligned images.

**Figure 4.**  (a) Schmid factor map of the region of interest and (b) maximum shear strain vs. Schmid factor at a nominal stress of 1280 MPa.

**Figure 5.**  (a) Maximum shear strain map with the grain boundaries overlaid and (b) maximum shear strain with respect to distance from grain boundary.

**Figures**

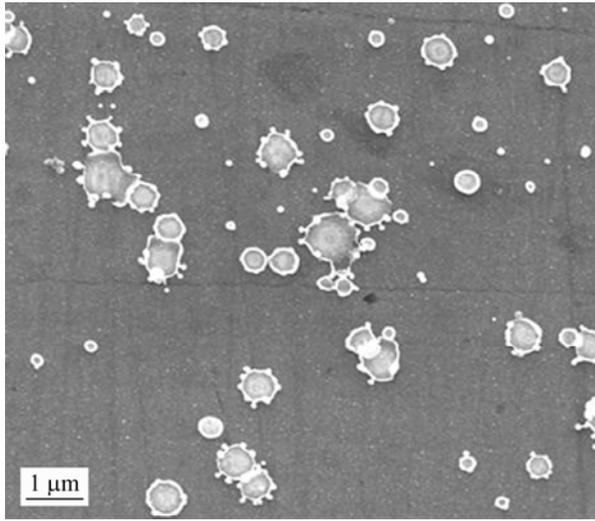

**Figure 1.** BSE images of nanoparticles on the surface of the tensile specimen.

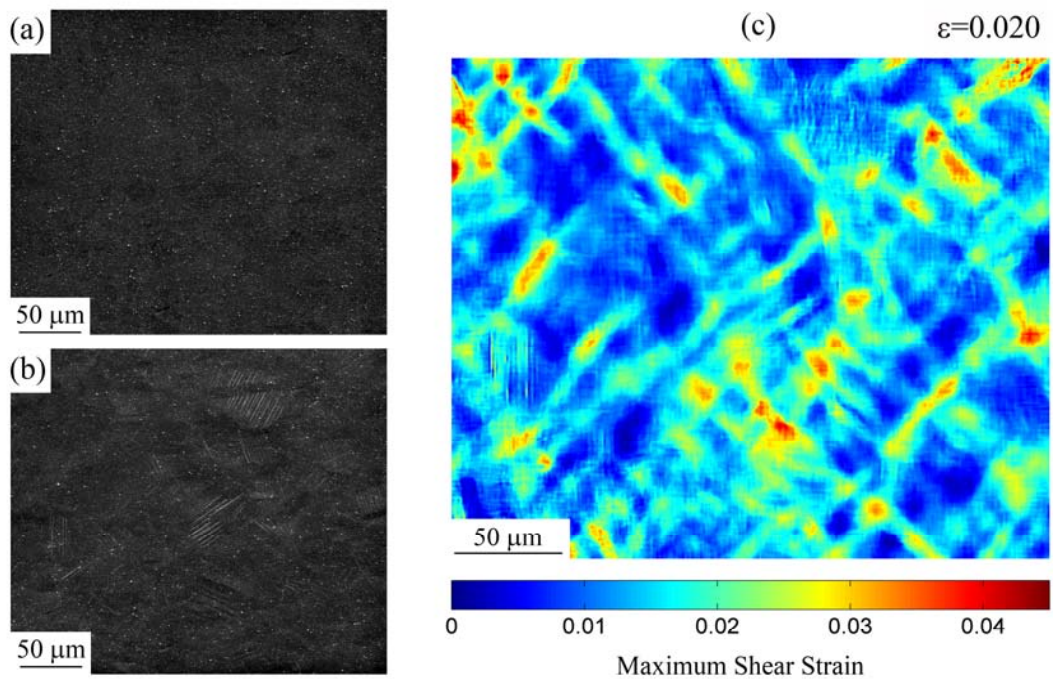

**Figure 2.** (a,b) SE images of René 88DT with nanoparticles (bright spots throughout image) at no load and after yield. (c) Digital image correlation was used to generate the maximum shear strains within the 274 μm by 230 μm region of interest.

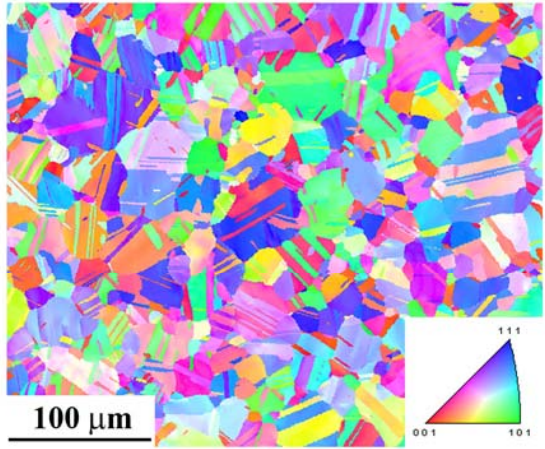
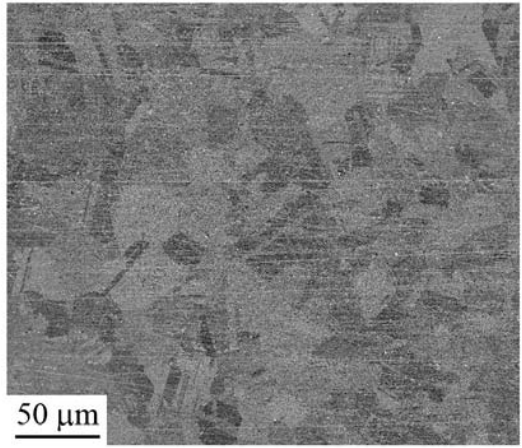
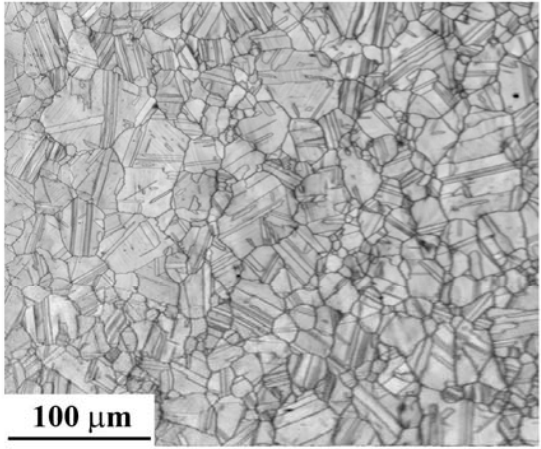
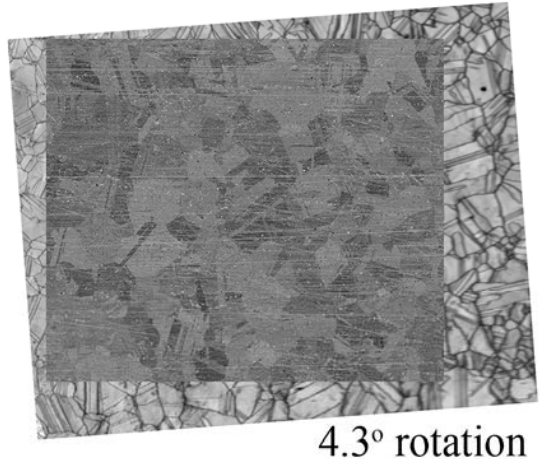

**Figure 3.** Image alignment process: (a) Inverse pole figure showing crystal orientations from EBSD scan, (b) BSE image of René 88DT prior to deformation, (c) IQ map of microstructure, and (d) aligned images.

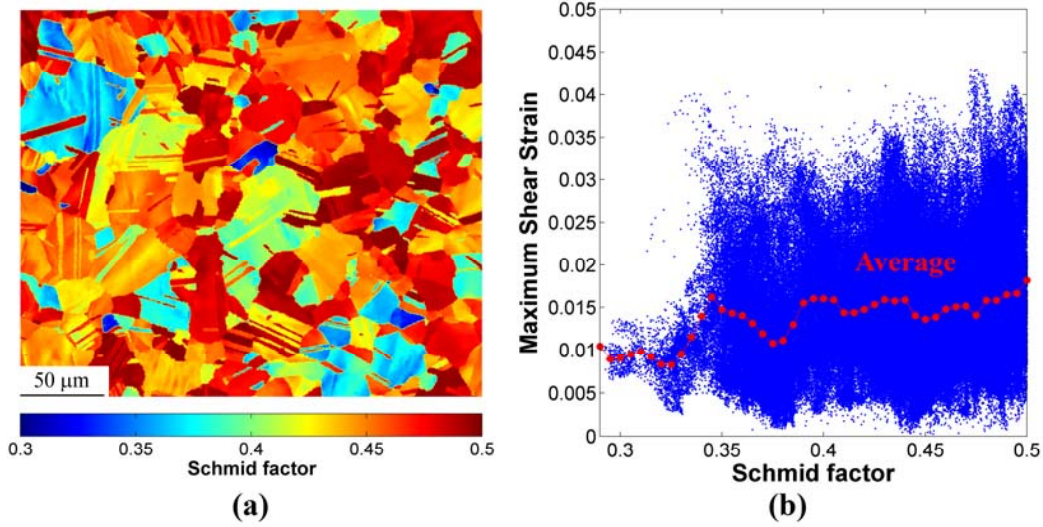

**Figure 4.** (a) Schmid factor map of the region of interest and (b) maximum shear strain vs. Schmid factor at a nominal stress of 1280 MPa.

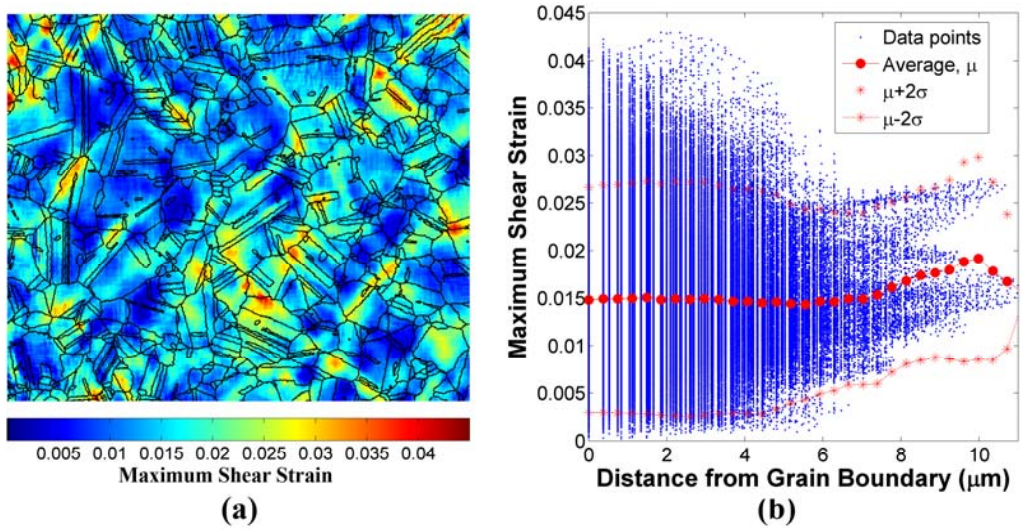

**Figure 5.** (a) Maximum shear strain map with the grain boundaries overlaid and (b) maximum shear strain with respect to distance from grain boundary.